\documentclass[twocolumn,showpacs,preprintnumbers,amsmath,amssymb,nofootinbib,APS]{revtex4}

\usepackage{graphicx}
\usepackage{dcolumn}
\usepackage{bm}

\begin{document}

\title{ {\bf The gravity lagrangian according to solar system experiments}}
\author{Gonzalo J. Olmo}\email{gonzalo.olmo@uv.es}
\affiliation{ {\footnotesize Departamento de F\'{\i}sica Te\'orica and
    IFIC, Centro Mixto Universidad de
    Valencia-CSIC \\
 Universidad de Valencia, Burjassot-46100, Valencia, Spain}\\
    {\footnotesize  Physics Department, University of
Wisconsin-Milwaukee,Milwaukee, WI 53201, USA }}
\date{May 19th, 2005}


\begin{abstract}
In this work we show that the gravity lagrangian $f(R)$ at
relatively low curvatures in both metric and Palatini formalisms
is a bounded function that can only depart from the linearity
within the limits defined by well known functions. We obtain those
functions by analyzing a set of inequalities that any $f(R)$
theory must satisfy in order to be compatible with laboratory and
solar system observational constraints. This result implies that
the recently suggested $f(R)$ gravity theories with nonlinear
terms that dominate at low curvatures are incompatible with
observations and, therefore, cannot represent a valid mechanism to
justify the cosmic speed-up.
\end{abstract}

\maketitle

Observations carried out in the last few years, indicate that the
universe is undergoing a period of accelerated expansion
\cite{Tonry03-Knop03}. In the context of this cosmic speed-up,
modified theories of gravity in which the gravity lagrangian is a
nonlinear function of the scalar curvature
\begin{equation}\label{eq:f}
S=\frac{1}{2\kappa^2}\int d^4 x\sqrt{-g}f(R)+S_m[g_{\mu
\nu},\psi_m]
\end{equation}
have become object of recent interest. These theories are commonly
referred to as $f(R)$ gravities. Motivated by the fact that the
addition of positive powers of the scalar curvature to the
Hilbert-Einstein lagrangian may give rise to early-time
inflationary periods, modifications of the lagrangian that become
dominant at low curvatures have been suggested to justify the
observed late-time cosmic acceleration \cite{CDTT} (see also
\cite{CAPo}). The aim of these theories is to describe the cosmic
acceleration as an effect of the gravitational dynamics itself
rather than as due to the existence of sources of dark energy. In
addition to the standard formulation of $f(R)$ gravities, where
the metric is the only gravitational field, it was pointed out in
\cite{VoL} that once nonlinear terms are introduced in the
lagrangian, an alternative and inequivalent formulation of these
theories is possible. If one considers metric and connection as
independent fields, i.e., that the connection in the gravity
lagrangian is not the usual Levi-Civita connection but is to be
determined by the equations of motion, then the resulting theory
is different from that defined in terms of the metric only. This
new approach, known as Palatini formalism, and the metric one have
been shown to lead to late-time cosmic acceleration for many
different $f(R)$ lagrangians. However, the attempts made so far to
unravel the form of the function $f(R)$ from the cosmological data
are far from being conclusive \cite{matching-data}.\\
In this work we study the constraints on $f(R)$ gravities imposed
by laboratory and solar system experiments. We find that the
lagrangian must be linear in $R$ and that the possible nonlinear
corrections are bounded by $R^2$. This implies that the cosmic
speed-up cannot be due to unexpected gravitational effects at low cosmic curvatures.\\

In order to confront the predictions of a given gravity theory
with experiment in the solar system, it is necessary to obtain its
weak field, slow motion (or post-Newtonian) limit. We will now
derive a scalar-tensor representation for $f(R)$ gravities that
will allow us to treat the metric and Palatini formulations in a
very similar manner and will simplify the computations of the
post-Newtonian metric. The following action, leads to the same
equations of motion as eq.(\ref{eq:f})
\begin{equation}\label{eq:f-AB}
S=\frac{1}{2\kappa^2}\int d^4
x\sqrt{-g}\left[f(A)+\left(R-A\right)B\right] +S_m
\end{equation}
where $A$ and $B$ are auxiliary fields. To fix our notation, we
will denote $R(g)$ the contraction $R(g)\equiv g^{\mu \nu }R_{\mu
\nu }$ with the Ricci tensor $R_{\mu \nu }$ given in terms of the
Levi-Civita connection of $g_{\mu \nu }$, and $R(\Gamma )$ the
contraction $R(\Gamma )\equiv g^{\mu \nu }R_{\mu \nu }$ with
$R_{\mu \nu }$ given in terms of a connection $\Gamma$ independent
of $g_{\mu \nu }$. Thus, the symbol $R$ in eq.(\ref{eq:f-AB}) must
be seen as $R=R(g)$ in the metric formalism, and as $R=R(\Gamma )$
in the Palatini formalism. It is easy to see that the functional
variation of eq.(\ref{eq:f-AB}) with respect to $A$ gives
$B-df(A)/dA=0$. This equation can be used to algebraically solve
for $A=A(B)$. Inserting this solution back into eq.(\ref{eq:f-AB})
and defining
\begin{equation}\label{eq:V}
V(B)=AB-f(A) \ ,
\end{equation}
then eq.(\ref{eq:f-AB}) in the metric formalism can be identified
with the case $\omega =0$ of the Brans-Dicke-like theories
\begin{equation} \label{eq:ST}
S=\frac{1}{2\kappa ^2 }\int d^4 x\sqrt{-{g}}\left[\phi
R(g)-\frac{\omega }{\phi}(\partial_\mu
\phi\partial^\mu\phi)-V(\phi) \right]+S_m
\end{equation}
where $\phi\equiv B $. Analogously, one can find a scalar-tensor
representation for the Palatini formulation. In this case, the
equations of motion for the connection lead to
\begin{equation}\label{eq:def-Gamma}
\Gamma^\alpha_{\beta \gamma }=\frac{t^{\alpha \lambda
}}{2}\left(\partial_\beta t_{\lambda \gamma }+\partial_\gamma
t_{\lambda \beta }-\partial_\lambda t_{\beta \gamma }\right)
\end{equation}
where the tensor $t_{\mu \nu }$ is defined as $t_{\mu \nu
}=Bg_{\mu \nu }$. This solution for the connection allows us to
write $R(\Gamma)$ in terms of the metric and the field $B$ as
follows
\begin{equation}
R(\Gamma )=R(g)+\frac{3}{2B}\partial_\lambda B\partial^\lambda
B-\frac{3}{B}\Box B
\end{equation}
Inserting this solution in eq.(\ref{eq:f-AB}) and discarding a
total divergence we get
\begin{equation}\label{eq:fp-AB}
S=\frac{1}{2\kappa ^2}\int d^4x
\sqrt{-{g}}\left[B{R}({g})+\frac{3}{2B}\partial_{\mu }
B\partial^{\mu }B -V(B)\right]+S_m
\end{equation}
which represents the case $\omega =-3/2$ of the theories defined
in eq.(\ref{eq:ST}). It is worth noting that when $V(B)$ is given,
the inverse problem of finding the lagrangian $f(R)$ is also
possible. From eq.(\ref{eq:V}) we see that
\begin{equation}\label{eq:dVdB}
\frac{dV(B)}{dB}=A
\end{equation}
Using this algebraic equation to solve for $B=B(A)$, the
lagrangian can be written, using again eq.(\ref{eq:V}), as
\begin{equation}\label{eq:f(R)}
f(A)=AB-V(B)
\end{equation}
This means that, with regard to the equations of motion, $f(R)$
gravities in metric and Palatini formalisms are equivalent to
$\omega =0$ and $\omega =-3/2$ Brans-Dicke-like theories respectively.\\

The identification of $f(R)$ gravities with particular cases of
Brans-Dicke-like theories should, in principle, allow us to
express their post-Newtonian limit in a compact form dependent on
the parameter $\omega $. However, this is only partially true. A
glance to the equation of motion for the scalar field defined in
eq.(\ref{eq:ST})
\begin{equation}\label{eq:Box}
 [3+2\omega]
{\Box}\phi+2V(\phi)-\phi V'(\phi)= \kappa ^2 {T}
\end{equation}
where $T\equiv g^{\mu \nu }T_{\mu \nu }$, indicates that the field
is a dynamical object for all $\omega $ but $\omega =-3/2$. Thus,
there exists a clear dynamical difference between a generic
$\omega=$ constant theory and the case $\omega =-3/2$, which
corresponds to the Palatini form of $f(R)$ gravities. As a
scalar-tensor theory, the case $\omega=-3/2$ seems to have been
almost avoided in the literature. In the original Brans-Dicke
theory, where the potential term was not present, this case was
obviously pathological (see \cite{NI} for a discussion of the
limit $\omega \to -3/2$). Moreover, it was found that in order to
get agreement between predictions and solar system experiments
$\omega $ should be large and positive, and little attention was
paid later, when non-trivial potentials were considered, to small
or negative values of $\omega $. on the other hand, for $\omega
\neq -3/2$ theories with $V\neq 0$ it is well known that if the
field has associated a large effective mass, the predictions of
the theory may agree with solar system experiments irrespective of
the value of $\omega $ \cite{WAG-SW}. However, such result assumes
that the field is near an extremum of its potential. For $\omega
=0$ ($f(R)$ in metric formalism) that condition, $dV/d\phi =0$,
cannot be imposed in general. This follows from the variation of
eq.(\ref{eq:f-AB}) with respect to $B$ and eq.(\ref{eq:dVdB}),
which lead to $dV/d\phi =R(g)$. Since the leading order of $R(g)$
at a given time coincides with the background cosmic curvature
$R_0\neq 0$, we cannot impose $dV/d\phi =0$ for $\omega =0$
theories. We are thus forced to compute the post-Newtonian limit
of this case without making any a priori assumption or
simplification about the behavior of the potential. In the case
$\omega =-3/2$ the relation between $\phi $ and $V(\phi )$ (see
eq.(\ref{eq:Box})) is even stronger and also forces us not to make
any assumption on $V(\phi)$. \\

We shall now sketch the basic steps to compute the post-Newtonian
metric of Brans-Dicke-like theories, which will be detailed
elsewhere. We generalize the results of the literature so as to
include all the terms that are relevant for our discussion. For
approximately static solutions, corresponding to masses such as
the Sun or Earth, to lowest order, we can drop the terms involving
time derivatives from the equations of motion. In a coordinate
system in free fall with respect to the surrounding cosmological
model, the metric can be expanded about its Minkowskian value as
$g_{\mu \nu }= \eta_{\mu \nu }+h_{\mu \nu }$.\\  For $\omega\neq
-3/2$, the field is dynamical and can be expanded as $\phi
=\phi_0+\varphi (t,x)$, where $\phi _0\equiv\phi (t_0)$ is the
asymptotic cosmic value, which is a slowly-varying function of the
cosmic time $t_0$, and $\varphi (t,x)$ represents the local
deviation from $\phi _0$. To lowest order $(T\approx-\rho )$, the
metric satisfies the following equations
\begin{eqnarray}\label{eq:1}
-\frac{1}{2}\nabla^2\left[h_{00}-\frac{\varphi }{\phi
_0}\right]&=&\frac{\kappa ^2\rho }{2\phi _0}-\frac{V_0}{2\phi
_0}\\
-\frac{1}{2}\nabla^2\left[h_{ij}+\delta _{ij}\frac{\varphi}{\phi
_0}\right]&=&\delta _{ij}\left[\frac{\kappa ^2\rho}{2\phi _0}
+\frac{V_0}{2\phi _0}\right]\label{eq:2}
\end{eqnarray}
where the gauge condition $h^\mu _{k,\mu }-\frac{1}{2}h^\mu _{\mu
,k}=\partial_k\varphi^{(2)} /\phi _0$ has been used. In
eliminating the zeroth-order terms in the field equation for
$\varphi$, corresponding to the cosmological solution for $\phi
_0$, we obtain
\begin{equation}\label{eq:3}
\left[\nabla^2-m_\varphi^2\right]\varphi^{(2)} (t,x)=-\frac{\kappa
^2\rho }{3+2\omega }
\end{equation}
where $m^2_\varphi$ is a slowly-varying function of the cosmic
time
\begin{equation} \label{eq:mass}
m^2_\varphi \equiv \frac{\phi _0 V''_0-V'_0}{3+2\omega }
\end{equation}
Solving eqs.(\ref{eq:1}), (\ref{eq:2}) and (\ref{eq:3}), the
metric becomes
\begin{eqnarray}\label{eq:h00-M}
h_{00}&=& 2G\frac{M_{\odot}}{r}+\frac{V_0}{6\phi _0}r^2\\
h_{ij}&=& \delta _{ij}\left[2\gamma
G\frac{M_{\odot}}{r}-\frac{V_0}{6\phi
_0}r^2\right]\label{eq:hij-M}
\end{eqnarray}
where $M_{\odot}=\int d^3x'\rho(t,x')$ and we have defined the
effective Newton's constant and the PPN parameter $\gamma $ as
\begin{eqnarray}\label{eq:G-st}
G&=&\frac{\kappa ^2}{8\pi \phi_0}\left(1+\frac{F(r)}{3+2\omega }\right)\\
\gamma &=&\frac{3+2\omega -F(r)}{3+2\omega +F(r)}\label{eq:g-st}
\end{eqnarray}
The function $F(r)$ is given by $F(r)=e^{-|m_\varphi| r}$ when
$m_\varphi ^2>0$ and by $F(r)=\cos |m_\varphi| r$ when $m_\varphi
^2<0$. \\

Let us now obtain the metric for $\omega=-3/2$ theories. In this
case no boundary conditions are needed for the scalar field, since
it satisfies an algebraic equation. Denoting by $\phi =\phi (T)$
the solution to eq.(\ref{eq:Box}) when $\omega=-3/2$, we construct
the quantity $\Omega(T) \equiv \log[\phi /\phi _0]$, where the
subindex now denotes vacuum value, $\phi _0\equiv\phi (T=0)$.
Using the gauge condition $h^\mu _{k,\mu }-\frac{1}{2}h^\mu _{\mu
,k}=\partial_k\Omega(T)$, the equations for the metric can be
written as
\begin{eqnarray}
-\frac{1}{2}\nabla^2\left[h_{00}-\Omega(T)\right]&=&
\frac{\kappa ^2\rho-V(\phi )}{2\phi}\\
-\frac{1}{2}\nabla^2\left[h_{ij}+\delta _{ij}\Omega(T)\right]&=&
\left[\frac{\kappa ^2\rho+V(\phi )}{2\phi}\right]\delta _{ij}
\end{eqnarray}
The solution to these equations are
\begin{eqnarray}\label{eq:h00-P}
h_{00}(t,x)&=& 2G\frac{M_{\odot}}{r}+\frac{V_0}{6\phi _0}r^2+\Omega(T)\\
h_{ij}(t,x)&=& \left[2\gamma G\frac{M_{\odot}}{r}-\frac{V_0}{6\phi
_0}r^2-\Omega(T )\right]\delta _{ij}\label{eq:hij-P}
\end{eqnarray}
where $M_\odot\equiv \phi _0\int d^3x' \rho (t,x')/\phi $ and we
have defined
\begin{eqnarray}\label{eq:G-Pal}
G&=&\frac{\kappa ^2}{8\pi \phi_0}\left(1+\frac{M_V}{M_\odot}\right)\\
\gamma &=&\frac{M_\odot-M_V}{M_\odot+M_V}\label{eq:g-Pal}
\end{eqnarray}
with $M_V\equiv\kappa ^{-2}\phi _0\int d^3x' [V_0/\phi
_0-V(\phi)/\phi]$.\\

The term $(V_0/\phi _0)r^2$ appearing in eqs.(\ref{eq:h00-M}) and
(\ref{eq:hij-M}) for $\omega \neq -3/2$ and in (\ref{eq:h00-P})
and (\ref{eq:hij-P}) for $\omega =-3/2$ is related to the scalar
energy density and acts in a manner similar to the effects of  a
cosmological constant. Thus, any viable theory must give a
negligible contribution of this type. Let us now focus on
$\omega=0$ theories. The oscillating solutions $F(r)=\cos
|m_\varphi| r$ are always unphysical. If $|m_\varphi| L\ll 1$ with
$L$ large compared to solar system scales (long-range
interaction), we find $\gamma \approx 1/2$, which is ruled out by
observations ($\gamma_{obs} \approx 1$, see \cite{WIL-liv}). If
$|m_\varphi| L\gg 1$ (short-range), Newton's constant strongly
oscillates in space and the Newtonian limit is dramatically
modified. Thus, only the damped case $F(r)=e^{-|m_\varphi| r}$
with
\begin{equation}\label{eq:ratio-M}
m_\varphi ^2L^2\gg 1
\end{equation}
is physically acceptable. This condition means that the scalar
interaction range $l_\varphi =m_\varphi ^{-1}$ must be shorter
than any currently accessible experimental length $L$.\\
In the case $\omega =-3/2$, we can learn about the dependence of
$\phi $ on $T$ by studying the behavior of $M_\odot$, $G$ and
$\gamma $. According to the definition of $M_\odot$ and $M_V$ (see
below eqs.(\ref{eq:hij-P}) and (\ref{eq:g-Pal})), it follows that
a body with Newtonian mass $M_N\equiv \int d^3x' \rho (t,x')$ may
yield different values of $M_\odot$, $G$ and $\gamma $ depending
on its internal structure and composition. In consequence, a given
amount of Newtonian mass could lead to gravitational fields of
different strengths and dynamical properties. We thus demand a
very weak dependence of $\phi $ on $T$ so as to guarantee that
$M_\odot$, $G$ and $\gamma $ are almost constant. This is
consistent with the requirement that the local term $\Omega (T)$
in eqs.(\ref{eq:h00-P}) and (\ref{eq:hij-P}) must be small
compared to unity. Since the contribution of $\Omega (T)$ to the
acceleration of a body is given in terms of its gradient, we must
demand that
\begin{equation}\label{eq:const-P}
\left| \frac{T (\partial \phi /\partial T)}{\phi }\right|\ll 1
\end{equation}
from $T=0$ up to nuclear densities. otherwise, a change in $\phi$
when going from its vacuum value $\phi _0$ outside atoms to its
value inside atoms would lead to observable effects in the motion
of macroscopic test bodies placed in the gravitational field
defined by eqs.(\ref{eq:h00-P}) and (\ref{eq:hij-P}). Since such
effects have not been observed, it follows that over a wide range
of densities $\phi \approx \phi _0+(\partial \phi /\partial
T)|_{T=0}T$, with $|\phi _0^{-1}(\partial \phi /\partial
T)|_{T=0}T|\ll 1$, must be a very good approximation. The weak
dependence on $T$ expressed by eq.(\ref{eq:const-P}) can be
written using eq.(\ref{eq:Box}) to evaluate $(\partial \phi
/\partial T)$ as
\begin{equation}\label{eq:ratio-P}
\left|\frac{(\kappa ^2\rho/\phi ) }{(\phi V''-V') }\right|\ll 1
\end{equation}
We can thus interpret this equation in a manner analogous to
eq.(\ref{eq:ratio-M}), as the ratio of a length associated to the
matter density, $\mathcal{L}^{2}(\rho )=[\kappa ^2\rho /\phi
_0]^{-1}$, over a length associated to the scalar field, $l_\phi
^2=m_\phi
^{-2}\equiv [(\phi V''-V')\phi /\phi _0]^{-1}$.\\

The constraint on $\omega =0$ theories given in
eq.(\ref{eq:ratio-M}) can be rewritten in terms of the lagrangian
$f(R)$ as follows
\begin{equation}\label{eq:const-Mf}
R_0\left[\frac{f'(R_0)}{R_0f''(R_0)}-1\right]L^2\gg 1
\end{equation}
where $R_0$ represents the current cosmic scalar curvature. Note
that since $\phi \equiv f'>0$ to have a well posed theory, it
follows that $f''$ must be small and positive in order to satisfy
eq.(\ref{eq:const-Mf}). In consequence, any theory with $f''<0$
leads to an ill defined post-Newtonian limit ($m_\varphi ^2<0$).
This is the case, for instance, of the Carroll et al. model
\cite{CDTT}. We shall now demand that the interaction range of the
scalar field remains as short as today or decreases with time so
as to avoid dramatic modifications of the gravitational dynamics
in post-Newtonian systems with the cosmic expansion. This can be
implemented imposing
\begin{equation}\label{eq:difeq-0}
\left[\frac{f'(R)}{Rf''(R)}-1\right]\ge \frac{1}{l^2R}
\end{equation}
as $R\to 0$, where $l^2\ll L^2$ represents a bound to the current
interaction range of the scalar field. Manipulating this
inequality we obtain
\begin{equation}\label{eq:difeq-1}
\frac{d\log[f'(R)]}{dR}\le \frac{l^2}{1+l^2R}
\end{equation}
which can be integrated twice to give
\begin{equation}\label{eq:fR-0}
f(R)\le \alpha +\beta \left(R+\frac{l^2R^2}{2}\right)
\end{equation}
where $\beta >0$. Since $f'$ and $f''$ are positive, the
lagrangian is also bounded from below, say, $f(R)\ge \alpha $.
According to the cosmological data, $\alpha \equiv -2\Lambda$ must
be of order a cosmological constant $2\Lambda \sim 10^{-53}
$m$^2$. Without loss of generality, setting $\beta=1$ we find that
in order to satisfy the current solar system constraints, the
lagrangian in metric formalism must satisfy
\begin{equation}\label{eq:fR-1}
-2\Lambda \le f(R)\le R-2\Lambda+\frac{l^2R^2}{2}
\end{equation}
which is clearly incompatible with nonlinear terms growing at low
curvatures. \\

Let us consider now the Palatini case, $\omega =-3/2$. Written in
terms of the lagrangian $f(R)$, eq.(\ref{eq:ratio-P}) turns into
\begin{equation}\label{eq:f-const}
R\tilde{f}'(R)\left|\frac{\tilde{f}'(R)}{R\tilde{f}''(R)}-1\right|
\mathcal{L}^2(\rho )\gg 1
\end{equation}
where $\tilde{f}'\equiv f'/f'_0=\phi /\phi _0$. Since
$\mathcal{L}^2(\rho )\sim 1/\rho$ takes its smallest value for
ordinary matter at nuclear densities, it is reasonable to demand
that
\begin{equation}\label{eq:f-const1}
\left|\frac{\tilde{f}'(R)}{R\tilde{f}''(R)}-1\right|\ge
\frac{1}{l^2 R\tilde{f}'}
\end{equation}
where $l^2$ represents a length scale much smaller than
$\mathcal{L}^2(\rho )$ at those densities. Note that $l^2$
determines the scale over which the nonlinear corrections are
relevant. If $l^2=0$, which implies $f(R)=a +b R$, then
eq.(\ref{eq:f-const}) would be valid for all $\rho $. Furthermore,
if the nonlinear corrections were important at very low cosmic
densities, $l$ would be of order the radius of the universe and
the nonlinear terms would dominate at all scales, which would lead
to unacceptable predictions, as we pointed out above with regard
to $M_\odot$, $G$, $\gamma $ and $\Omega (T)$. A good example of
such pathological effects in the Palatini formalism was studied in
\cite{FLA}. Manipulating eq.(\ref{eq:f-const1}) we obtain for
$\tilde{f}''>0$
\begin{equation}\label{eq:f''+}
{f}\le a+\frac{l^2{R}^2}{2}+\frac{{R}}{2}\sqrt{1+(l^2{R})^2}+
\frac{1}{2l^2}\log[l^2{R}+\sqrt{1+(l^2{R})^2}]
\end{equation}
and for $\tilde{f}''<0$
\begin{equation}\label{eq:f''-}
{f}\ge a -\frac{l^2{R}^2}{2}+\frac{{R}}{2}\sqrt{1+(l^2{R})^2}+
\frac{1}{2l^2}\log[l^2{R}+\sqrt{1+(l^2{R})^2}]
\end{equation}
where the vacuum value $f_0'$ has been set to unity and $a$ can be
identified with $a \equiv -2\Lambda $. We see that, to leading
order in $l^2R$, the Palatini lagrangian is bounded by
\begin{equation}\label{eq:bound}
R-2\Lambda -\frac{l^2R^2}{2}\le f(R) \le R-2\Lambda
+\frac{l^2R^2}{2}
\end{equation}
According to eqs.(\ref{eq:fR-1}) and (\ref{eq:bound}), our
conclusion is clear, laboratory and solar system experiments
indicate that the gravity lagrangian is nearly linear in $R$, with
the possible nonlinearities bounded by quadratic terms. In
consequence, $f(R)$ lagrangians with nonlinear terms that grow at
low curvatures cannot represent a valid mechanism to justify the
cosmic accelerated expansion rate. Such theories lead to a
long-range scalar interaction incompatible with the experimental
tests. In the viable models the non-linearities represent a
short-range scalar interaction, whose effect in the late-time
cosmic dynamics reduces to that of  a cosmological constant and,
therefore, do not substantially modify the description provided by
General Relativity. To conclude, we want to remark that
eqs.(\ref{eq:fR-1}) and (\ref{eq:f''+})-(\ref{eq:f''-}) tell us
how the lagrangian $f(R)$ must be near the origin, $l^2R\ll 1$,
not far from it, $l^2R\gg 1$, where the post-Newtonian constraints
could not make sense. This fact constraints the possible $f(R)$
early-time inflationary models. In particular, $f(R)=R+aR^2$ in
metric formalism seems compatible with CMBR observations \cite{JH-HN}.\\
I thank Prof. Leonard Parker and Prof. Jos\'{e} Navarro-Salas for
their wise advice and insights. This work was supported by a
fellowship from the Regional Government of Valencia (Spain) and
the research grant BFM2002-04031-C02-01 from the M.E.C.(Spain).

\end{document}